\documentclass[12pt,a4paper]{article}%

\usepackage{amsmath,amssymb,amsfonts}
\usepackage{caption2}
\usepackage[]{hyperref}
\usepackage[pdftex]{color,graphicx}
\usepackage{multicol}

\numberwithin{equation}{section} 
\setcaptionmargin{1cm}
\newcommand{\hhref}[1]{\href{http://arxiv.org/abs/#1}{{\it arXiv:#1}}}

\begin{document}

\begin{titlepage}
$\quad$
\vskip 2.0cm
\begin{center}
{\huge \bf  Threshold corrections to hard \\[3mm] supersymmetric relations }  
\vskip 1.0cm {\large Pier Paolo Giardino$^{a}$ and Paolo Lodone$^{bc}$ } \\[1cm]
{\it $^a$ Dipartimento di Fisica dell'Universit\`a di Pisa and INFN, Italy    \\
$^b$   Institut de Th\'eorie des Ph\'enom\`enes Physiques, EPFL, Lausanne, Switzerland  \\
$^c$ Scuola Normale Superiore di Pisa and INFN, Italy    
} \\[5mm]
\vskip 1.0cm
\today
\end{center}

\begin{abstract}
We compute the one-loop threshold corrections to hard supersymmetric relations involving gauge and gaugino couplings for the case of a Split-SUSY-like spectrum with moderate splitting.
We show that these contributions are small, however in principle they will typically have to be taken into account if the heavy scalar sector is below 100 TeV and if one reaches, at future linear colliders, the precision that is necessary to perform this kind of measurements.
\end{abstract}
\end{titlepage}


\section{Introduction}

If good superpartner-candidates are discovered at the LHC, the most important target will be the precise measurement of their masses and interactions.
The verification of softly broken Supersymmetry will have in fact to pass through the verification of well-defined relations among the masses and the couplings of these new particles \cite{KEK-PREPRINT-93-146,hep-ph/9502260,Nojiri:1996xx}.
The necessity of performing these measurements would then be a primary motivation for a linear $e^+ e^-$-collider.

Moreover, as first noted in \cite{Nojiri:1996xx}, the radiative corrections to these supersymmetric relations are logarithmically sensitive to large mass differences between the superpartners. Thus, in a situation in which some of the superpartners are relatively heavy and out of the reach of direct production, the precise measurement of the properties of the lighter ones can provide important indirect informations on the heavy part of the spectrum.
These effects have been intensively studied in the past \cite{hep-ph/9501382,Pierce:1996xx,Cheng:1997sq,hep-ph/9706476,Katz:1998br,hep-ph/9803210,Kiyoura:1998yt}, sometimes with the somewhat misleading name of ``Super-oblique corrections''.
In fact the approximation of retaining only the oblique contributions is typically not justified, the vertex corrections being of equal importance.
It is convenient instead to discuss these effects in an effective Lagrangian approach,
as done in \cite{Giudice:2004tc,Bernal:2007uv} in the context of Split-Supersymmetry \cite{ArkaniHamed:2004fb,Giudice:2004tc} which is one of the best motivated frameworks with very large mass splittings between the superpartner masses\footnote{For a recent discussion of a possible relatively split spectrum compatible instead with naturalness see \cite{Barbieri:2010pd}.}.

Let us focus on the Split-SUSY case, and recall the low energy spectrum contains the SM particles with one Higgs doublet $H$, the Higgsinos $\tilde{H}_{u,d}$ and the gauginos $\tilde{g},\tilde{W},\tilde{B}$, with typical mass $m_{light}$.
In the conventions of \cite{Giudice:2004tc} the most general renormalizable effective Lagrangian, with matter parity and the particle content assumed above, is given by:
\begin{eqnarray}
\mathcal{L} &=& m^2 H^\dagger H - \frac{\lambda}{2}(H^\dagger H)^2 - \left[ h_{ij}^u \overline{q}_j u_i \epsilon H^* + h_{ij}^d \overline{q}_j d_i H + h_{ij}^e \overline{\ell}_j e_i H  \right. \label{eq:effLagr} \\
&&   + \frac{M_3}{2} \tilde{g}^A \tilde{g}^A  + \frac{M_2}{2} \tilde{W}^a \tilde{W}^a + \frac{M_1}{2} \tilde{B} \tilde{B} + \mu \tilde{H}_u^T \epsilon \tilde{H}_d \nonumber \\
&&\left. + \frac{H^\dagger}{\sqrt{2}}(\tilde{g}_u \sigma^a \tilde{W}^a + \tilde{g}'_u \tilde{B}) \tilde{H}_u  + \frac{H^T \epsilon}{\sqrt{2}}(-\tilde{g}_d \sigma^a \tilde{W}^a + \tilde{g}'_d \tilde{B}) \tilde{H}_d + \mbox{h.c.} \right] \nonumber
\end{eqnarray}
where $\epsilon=i\sigma^2$ and the gauge-invariant kinetic terms have been omitted.
If one matches at tree level the couplings $\lambda, h^{u,d,e},\tilde{g}_{u,d},\tilde{g}^{\prime}_{u,d}$ with the ones of the full MSSM Lagrangian at the scale $\tilde{m}$ at which the heavy fields are decoupled, then one finds:
\begin{equation} \label{eq:susyrel:1}
\lambda(\tilde{m})=\frac{g^2(\tilde{m}) +g^{\prime 2}(\tilde{m})}{4}\cos^2 2\beta
\end{equation}
\begin{eqnarray}
h^u_{ij}(\tilde{m}) = \lambda^{u *}_{ij}(\tilde{m})  \sin\beta&&h^{d,e}_{ij}(\tilde{m})  = \lambda^{d,e *}_{ij}(\tilde{m})  \cos\beta  \\ 
\tilde{g}_u(\tilde{m})  =  g(\tilde{m})  \sin\beta && \tilde{g}_d(\tilde{m})  = g(\tilde{m})  \cos\beta \\  
\tilde{g}'_u(\tilde{m}) =  g'(\tilde{m})  \sin\beta  && \tilde{g}'_d(\tilde{m})  = g'(\tilde{m})  \cos\beta  \label{eq:susyrel:3}
\end{eqnarray}
where $\lambda^{u,d,e}$ and $g,g'$ are the Yukawa and gauge couplings.
When the heavy fields are decoupled, the gauge couplings $g,g'$ and the gaugino ones $\tilde{g}_{u,d},\tilde{g}^{\prime}_{u,d}$ run differently down to low energy, and this results in a calculable mismatch with respect to the supersymmetric relations (\ref{eq:susyrel:1})-(\ref{eq:susyrel:3}).
In fact, if the scale $\tilde{m}$ is very far away, as it can be in Split Supersymmetry, then the large logarithms of the ratio $\tilde{m}/m_{light}$ need to be resummed. To this end, in \cite{Giudice:2004tc} the relevant RGEs are reported, under which the above couplings evolve down to low energy. These RGEs can be obtained from the general results \cite{Machacek:1983tz}\cite{Machacek:1983fi}\cite{Machacek:1984zw} and can be seen to match the supersymmetric case \cite{Martin:1993zk} once the heavy superpartners are reintroduced.
On the other hand, besides the logarithmically enhanced contributions there are also finite threshold terms coming from the matching at the scale at which the heavy superpartners are integrated out.
These kind of terms are usually taken into account only when one considers the two loop running (see e.g. \cite{arXiv:1108.6077}), however as already noticed in \cite{Cheng:1997sq,Katz:1998br} if the logarithms are not so large then the various threshold contributions may become numerically important.
In fact in \cite{Bernal:2007uv} the one loop threshold correction of $O(\lambda_t^4)$ to the quartic coupling $\lambda$ is included, but the threshold corrections to the relations between the gauge and gaugino couplings are still missing in the literature.
Our scope is to compute, in the context of Split-SUSY, all these finite terms which have been so far computed only partially or in particular cases. Thus the situation in which ``Supersymmetry is Split but not so much'' will also be covered.

\section{Results}

Details about the computation are reported in the Appendix \ref{app:matching}.
The final result in the $\overline{DR}$ renormalization scheme is, keeping only the top Yukawa coupling:
\begin{eqnarray}
\left. \frac{\tilde{g}'_u}{g' \sin \beta} \right|_{{\mu}}  &=& 1 + \frac{\log \frac{\tilde{m}}{\mu}}{(4\pi)^2} \left[ - \frac{9}{4} \cos^2 \beta \, g^{2}  + \left( 7 - \frac{3}{4} \cos^2 \beta \right)g^{\prime 2} - 3 h_t^2  \right]    \label{eq:gprime:up}  \\
&& + \frac{1}{(4\pi)^2} \left[  \frac{21}{16} \cos^2 \beta \, g^{2}  + \left( -\frac{21}{8} + \frac{7}{16} \cos^2 \beta \right)g^{\prime 2} + \left(  \frac{3}{4\sin^2 \beta} + \frac{3}{ 2}  \right) h_t^2 \right] \nonumber \\
\left. \frac{\tilde{g}_u}{g \sin \beta} \right|_{{\mu}} &=& 1 +  \frac{\log \frac{\tilde{m}}{\mu}}{(4\pi)^2} \left[ \left( \frac{13}{3} + \frac{7}{4} \cos^2 \beta \right)g^2 - \frac{3}{4} \cos^2 \beta \, g^{\prime 2} - 3 h_t^2  \right]  \label{eq:g:up} \\
&& + \frac{1}{(4\pi)^2} \left[ \left( -\frac{13}{8} - \frac{11}{16} \cos^2 \beta \right)g^2 + \frac{7}{16} \cos^2 \beta \, g^{\prime 2} + \left(  \frac{3}{4\sin^2 \beta} + \frac{3}{2}  \right) h_t^2  \right] \,   \nonumber
\end{eqnarray}
\begin{eqnarray}
 \left. \frac{\tilde{g}'_d}{g' \cos \beta} \right|_{{\mu}}  &=& 1 + \frac{\log \frac{\tilde{m}}{\mu}}{(4\pi)^2} \left[ - \frac{9}{4} \sin^2 \beta \, g^{2}  + \left( 7 - \frac{3}{4} \sin^2 \beta \right)g^{\prime 2} - 3 h_t^2  \right]    \label{eq:gprime:down}  \\
&& + \frac{1}{(4\pi)^2} \left[  \frac{21}{16} \sin^2 \beta \, g^{2}  + \left( -\frac{21}{8} + \frac{7}{16} \sin^2 \beta \right)g^{\prime 2} + \frac{3}{2} h_t^2 \right] \nonumber \\
\left. \frac{\tilde{g}_d}{g \cos \beta} \right|_{{\mu}} &=& 1 +  \frac{\log \frac{\tilde{m}}{\mu}}{(4\pi)^2} \left[ \left( \frac{13}{3} + \frac{7}{4} \sin^2 \beta \right)g^2 - \frac{3}{4} \sin^2 \beta \, g^{\prime 2} - 3 h_t^2  \right]  \label{eq:g:down} \\
&& + \frac{1}{(4\pi)^2} \left[ \left( -\frac{13}{8} - \frac{11}{16} \sin^2 \beta \right)g^2 + \frac{7}{16} \sin^2 \beta \, g^{\prime 2}   + \frac{3}{2}   h_t^2  \right] \,  , \nonumber
\end{eqnarray}
where $\mu$ is the renormalization scale that will be typically chosen to be $\mu \sim m_{light}$, and all the terms suppressed by powers of the heavy masses have been neglected.
Notice that neglecting the finite threshold terms one recovers the equations (26) and (27) of \cite{Giudice:2004tc}.

In the $\overline{MS}$ scheme one has additional threshold corrections due to the fact that the regularization does not respect Supersymmetry. These additional terms are reported for completeness in Appendix \ref{app:MSandDR}.

Equations (\ref{eq:gprime:up})-(\ref{eq:g:down}) are appropriate assuming that the heavy superpartners have a common mass $\tilde{m}$. The precise form of the heavy spectrum can introduce additional threshold corrections that can be easily derived from the expressions that we report in the Appendix \ref{app:matching} for generic heavy masses.

\begin{figure}
\begin{center}
\begin{tabular}{cc}
\includegraphics[width=0.44\textwidth]{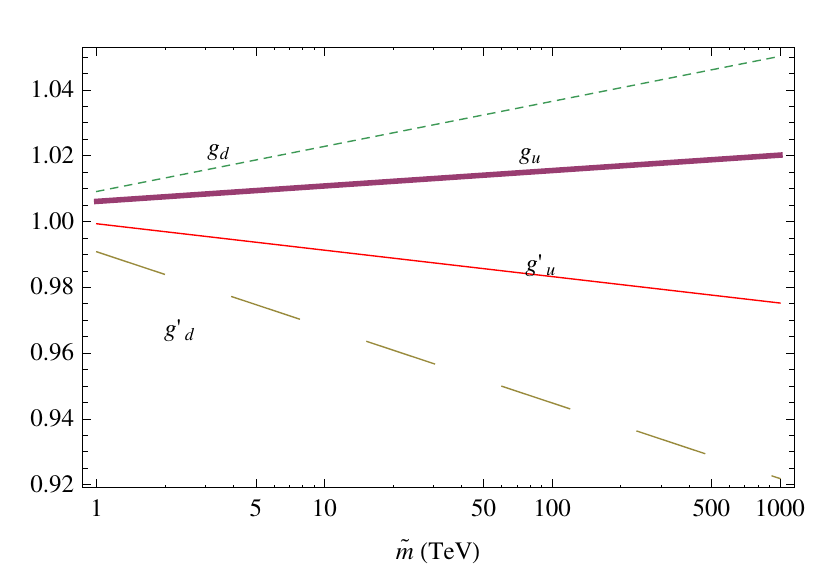} &
\includegraphics[width=0.44\textwidth]{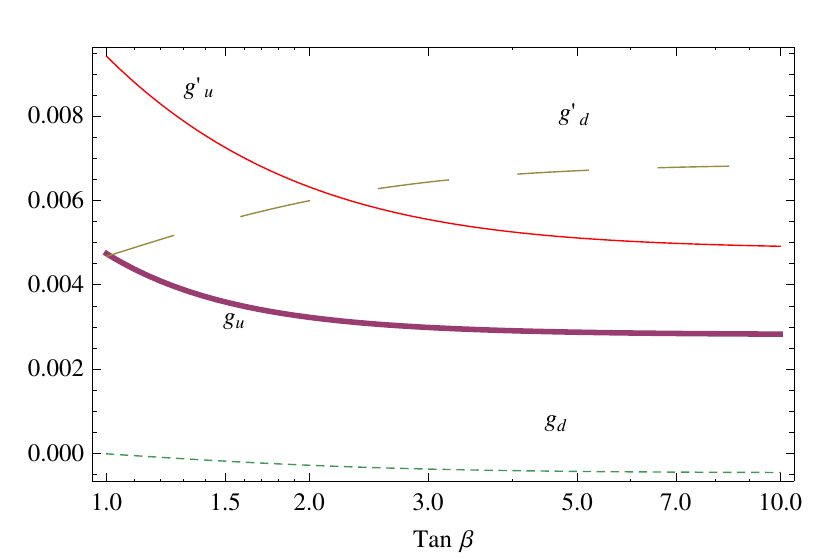}
\end{tabular}
\caption{\small {\it Left: plot of the ratios (\ref{eq:gprime:up})-(\ref{eq:g:down}) for $\tan\beta=10$ and $\mu=200$ GeV. Right: threshold contributions only, to the same ratios, as a function of $\tan\beta$. }}
\label{fig:figure}
\end{center}
\end{figure}

The typical size of the deviation from $1$ of the ratios (\ref{eq:gprime:up})-(\ref{eq:g:down}) is of order of a few percent for $\tilde{m}\lesssim 10^3$ TeV, as graphically shown in Figure \ref{fig:figure} Left.
At the same time the size of the threshold corrections that we computed is at most of order 1\%, as shown in Figure \ref{fig:figure} Right.
We conclude that the impact of these threshold corrections on the supersymmetric relations among gauge and gaugino couplings is negligible, for practical purposes, until a precision at the percent level is reached in the measurement of these observables.
Notice however that it will be compulsory to reach such a high precision, in order to start probing the radiative corrections to these relations, if $\tilde{m}/m_{light}$ is not much larger than $10^3$.
In this last case the threshold corrections of the type that we computed would need to be taken into account.

\section*{Acknowledgments}

We thank Alessandro Strumia and Gian Giudice for discussions and encouragement. 
This work has been supported in part by the EU ITN ``Unification in the LHC Era'', contract PITN-GA-2009-237920 (UNILHC).

\appendix
\section{Matching procedure} \label{app:matching}

\begin{figure}[tbh]
\begin{center}
\includegraphics[width=0.80\textwidth]{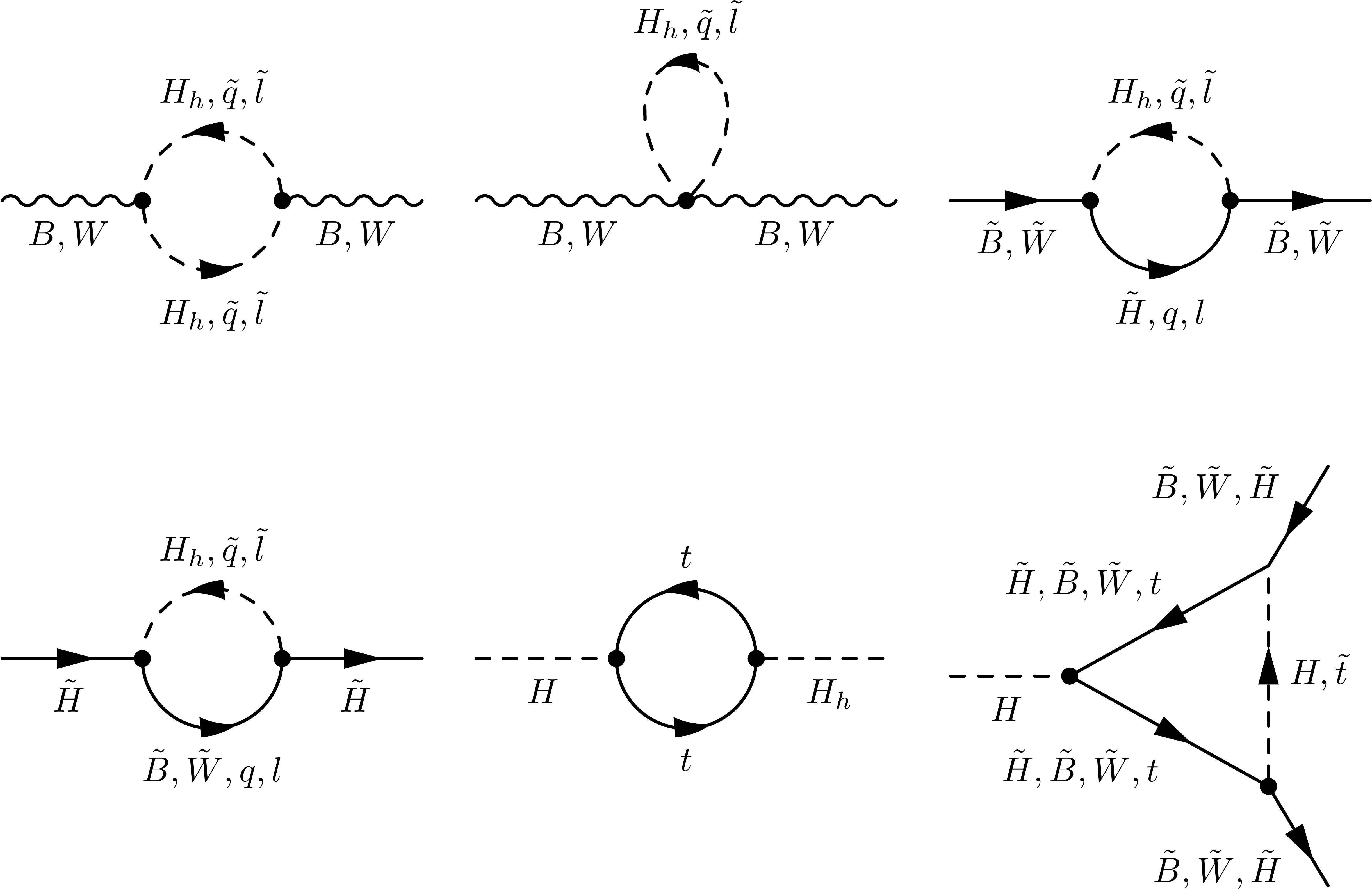}
\caption{\small {\it Relevant diagrams for the matching described in Appendix \ref{app:matching}, see text. All the other contributions are either suppressed or they cancel one another.}}
\label{fig:Feyn}
\end{center}
\end{figure}

To obtain the results (\ref{eq:gprime:up})-(\ref{eq:g:down}) we match at one loop  the couplings $g_{(eff)}$s of the effective theory (\ref{eq:effLagr}) with the couplings $g_{(full)}$s of the full theory at a scale $\overline{\mu} \sim \tilde{m}$. To this end it is sufficient to consider the diagrams that involve some heavy virtual particle, since all the other diagrams give the same contribution in the two cases. 
The relevant Feynman diagrams are then reported in Figure \ref{fig:Feyn}.
The result for the gauge couplings is, in $\overline{DR}$ after the minimal subtraction:
\begin{eqnarray}
g'_{(full)}(\overline{\mu}) &=& g'_{(eff)}(\overline{\mu}) \times \left\{ 1 + \frac{g^{\prime 2}}{(4\pi)^2} \, \frac{7}{2} \log \frac{\overline{\mu}}{\tilde{m}_1}  \right\}   \label{eq:gprimeMSSMeff}  \\
g_{(full)}(\overline{\mu}) &=& g_{(eff)}(\overline{\mu}) \times \left\{ 1 + \frac{g^{ 2}}{(4\pi)^2} \, \frac{13}{6} \log \frac{\overline{\mu}}{\tilde{m}_2}  \right\}        
\end{eqnarray}
where all comes from self energy diagrams, and:
\begin{eqnarray}
\tilde{m}_1 =  m_{H_h}^{1/21} \prod_{i=1}^3 m_{\tilde{q}_i}^{11/63} \prod_{j=1}^3 m_{\tilde{\ell}_j}^{1/7} 
& , &
\tilde{m}_2 = m_{H_h}^{1/13} \prod_{i=1}^3 m_{\tilde{q}_i}^{3/13} \prod_{j=1}^3 m_{\tilde{\ell}_j}^{1/13}  \, .
\end{eqnarray}
In the above expressions ${H_h}$ is the heavy Higgs scalar doublet and the products are over squark and slepton generations, that are assumed to have a common mass (further generalization is straightforward).

On the other hand for the gaugino couplings we find:
\begin{eqnarray}
\tilde{g}'_{u \, (full)}(\overline{\mu}) &=& \tilde{g}'_{u \, (eff)}(\overline{\mu}) \times \left\{ 1 + \frac{1}{(4\pi)^2}\, \left[   \frac{21}{2} \left(  \log \frac{\overline{\mu}}{\tilde{m}_1} + \frac{1}{4} \right)g^{\prime 2} + 3 \lambda_t^2 \left( \log \frac{\overline{\mu}}{{m}_{\tilde{q}_3}}  + \frac{1}{4} \right)  \right. \right. \\
&& \left. +  3 \lambda_t^2 \cos^2 \beta \left( \log \frac{\overline{\mu}}{{m}_{H_h}}  + \frac{1}{2} \right) + \left( \frac{g^{\prime 2}}{4} + \frac{3 g^2}{4} \right)\cos^2\beta \left( \log\frac{\overline{\mu}}{{m}_{H_h}} + \frac{1}{4} \right)  \right]  \nonumber \\
&& \left. + \frac{1}{(4\pi)^2} \, \left[    - 4 \left( \frac{g^{\prime 2}}{4} + \frac{3 g^2}{4} \right)\cos^2\beta \left( \log\frac{\overline{\mu}}{{m}_{H_h}} + \frac{1}{2} \right)   -6 \lambda_t^2 \left( \log \frac{\overline{\mu}}{{m}_{\tilde{q}_3}}  + \frac{1}{2} \right)    \right] \right\}      \nonumber  \\
\tilde{g}_{u \, (full)}(\overline{\mu}) &=& \tilde{g}_{u \, (eff)}(\overline{\mu}) \times \left\{  1 + \frac{1}{(4\pi)^2} \,\left[   \frac{13}{2} \left(  \log \frac{\overline{\mu}}{\tilde{m}_2} + \frac{1}{4} \right)g^{2} + 3 \lambda_t^2 \left( \log \frac{\overline{\mu}}{{m}_{\tilde{q}_3}}  + \frac{1}{4} \right)  \right. \right. \\
&& \left. +  3 \lambda_t^2 \cos^2 \beta \left( \log \frac{\overline{\mu}}{{m}_{H_h}}  + \frac{1}{2} \right) + \left( \frac{g^{\prime 2}}{4} + \frac{3 g^2}{4} \right)\cos^2\beta \left( \log\frac{\overline{\mu}}{{m}_{H_h}} + \frac{1}{4} \right)  \right]  \nonumber \\
&& \left. + \frac{1}{(4\pi)^2} \, \left[    - 4 \left( \frac{g^{\prime 2}}{4} - \frac{ g^2}{4} \right)\cos^2\beta \left( \log\frac{\overline{\mu}}{{m}_{H_h}} + \frac{1}{2} \right)   -6 \lambda_t^2 \left( \log \frac{\overline{\mu}}{{m}_{\tilde{q}_3}}  + \frac{1}{2} \right)    \right] \right\}      \nonumber 
\end{eqnarray}
\begin{eqnarray}
\tilde{g}'_{d \, (full)}(\overline{\mu}) &=& \tilde{g}'_{d \, (eff)}(\overline{\mu}) \times \left\{ 1 + \frac{1}{(4\pi)^2}\, \left[   \frac{21}{2} \left(  \log \frac{\overline{\mu}}{\tilde{m}_1} + \frac{1}{4} \right)g^{\prime 2}   \right. \right. \\
&& \left. -  3 \lambda_t^2 \sin^2 \beta \left( \log \frac{\overline{\mu}}{{m}_{H_h}}  + \frac{1}{2} \right)  + \left( \frac{g^{\prime 2}}{4} + \frac{3 g^2}{4} \right)\sin^2\beta \left( \log\frac{\overline{\mu}}{{m}_{H_h}} + \frac{1}{4} \right)  \right]  \nonumber \\
&& \left. + \frac{1}{(4\pi)^2} \, \left[    - 4 \left( \frac{g^{\prime 2}}{4} + \frac{3 g^2}{4} \right)\sin^2\beta \left( \log\frac{\overline{\mu}}{{m}_{H_h}} + \frac{1}{2} \right)     \right] \right\}      \nonumber  \\
\tilde{g}_{d \, (full)}(\overline{\mu}) &=& \tilde{g}_{d \, (eff)}(\overline{\mu}) \times \left\{  1 + \frac{1}{(4\pi)^2} \,\left[   \frac{13}{2} \left(  \log \frac{\overline{\mu}}{\tilde{m}_2} + \frac{1}{4} \right)g^{2}  \right. \right. \label{eq:gtildedMSSMsusy} \\
&& \left. -  3 \lambda_t^2 \sin^2 \beta \left( \log \frac{\overline{\mu}}{{m}_{H_h}}  + \frac{1}{2} \right)   + \left( \frac{g^{\prime 2}}{4} + \frac{3 g^2}{4} \right)\sin^2\beta \left( \log\frac{\overline{\mu}}{{m}_{H_h}} + \frac{1}{4} \right)  \right]  \nonumber \\
&& \left. + \frac{1}{(4\pi)^2} \, \left[    - 4 \left( \frac{g^{\prime 2}}{4} - \frac{ g^2}{4} \right)\sin^2\beta \left( \log\frac{\overline{\mu}}{{m}_{H_h}} + \frac{1}{2} \right)     \right] \right\}  \, ,    \nonumber 
\end{eqnarray}
where $h_t = \lambda_t \sin\beta$.
In all the above expressions the first part comes from self energy diagrams, the second part from vertex diagrams, and all the terms suppressed by powers of the heavy masses have been neglected.

To find the relation between the $g_{(eff)}$s$(\overline{\mu})$, we finally use the fact that the full theory is supersymmetric above $m_{light}$, and thus if we use a scheme that respects Supersymmetry (like $\overline{DR}$) it will be:
\begin{equation}
g'_{(full)}(\overline{\mu}) \sin \beta = \tilde{g}'_{u \, (full)} (\overline{\mu})  \quad , \quad
g_{(full)}(\overline{\mu}) \sin \beta = \tilde{g}_{u \, (full)} (\overline{\mu})  \, , \label{eq:MSSMsusy}
\end{equation}
and analogously for $\tilde{g}'_{d}$ and $\tilde{g}_{d}$.
Notice that the fact that we are in $\overline{DR}$ and not in $\overline{MS}$ comes into play now, in equation (\ref{eq:MSSMsusy}), in which using the $\overline{MS}$ we would have instead the additional terms reported in equations (\ref{eq:gprime:up:MSDR})-(\ref{eq:g:down:MSDR}) below.
On the other hand equations (\ref{eq:gprimeMSSMeff})-(\ref{eq:gtildedMSSMsusy}) are exactly the same in $\overline{MS}$ since there are only scalars and fermions in the loops.

From equations (\ref{eq:gprimeMSSMeff})-(\ref{eq:MSSMsusy}) with a common heavy mass $\tilde{m}$ one obtains (\ref{eq:gprime:up})-(\ref{eq:g:down}) at the scale $\overline{\mu} \sim \tilde{m}$. But now we notice that with this procedure we obtained the full one loop logarithmic dependence on the energy scale, thus in practice we can replace $\overline{\mu}$ with $\mu \sim m_{light}$.

\section{Relation between $\overline{DR}$ and $\overline{MS}$} \label{app:MSandDR}

For completeness we report the additional threshold terms that are present in the $\overline{MS}$ scheme, as in the Appendix of \cite{Bernal:2007uv}:
\begin{eqnarray}
\left. \frac{\tilde{g}'_u}{g' \sin \beta} \right|_{\overline{MS}}  &=& \left. \frac{\tilde{g}'_u}{g' \sin \beta} \right|_{\overline{DR}} + \frac{1}{(4\pi)^2} \left[ \frac{3}{8}g^2 + \frac{1}{8}g^{\prime 2}   \right]    \label{eq:gprime:up:MSDR}  \\
\left. \frac{\tilde{g}_u}{g \sin \beta} \right|_{\overline{MS}} &=& \left. \frac{\tilde{g}_u}{g \sin \beta} \right|_{\overline{DR}} +  \frac{1}{(4\pi)^2} \left[ \frac{23 }{24}g^2 - \frac{1}{8 }g^{\prime 2} \right]  \label{eq:g:up:MSDR}
\end{eqnarray}
\begin{eqnarray}
 \left. \frac{\tilde{g}'_d}{g' \cos \beta} \right|_{\overline{MS}}  &=&  \left. \frac{\tilde{g}'_d}{g' \cos \beta} \right|_{\overline{DR}} + \frac{1}{(4\pi)^2} \left[ \frac{3}{8}g^2 + \frac{1}{8}g^{\prime 2} \right]    \label{eq:gprime:down:MSDR}  \\
\left. \frac{\tilde{g}_d}{g \cos \beta} \right|_{\overline{MS}} &=& \left. \frac{\tilde{g}_d}{g \cos \beta} \right|_{\overline{DR}} +  \frac{1}{(4\pi)^2} \left[ \frac{23 }{24}g^2 - \frac{1}{8 }g^{\prime 2} \right] \, .  \label{eq:g:down:MSDR}
\end{eqnarray}


\vspace{0.3cm}

\begin{multicols}{2}

\end{multicols}

\end{document}